\pdfoutput=1
\documentclass[a4paper,11pt]{article}
\usepackage[utf8]{inputenc}

\usepackage{jcappub}
\usepackage{array}
\usepackage{hyperref}
\usepackage{graphicx}
\usepackage{color}
\usepackage{tensor}
\usepackage{cancel,xcolor}

\newcommand{\be}{\begin{equation}}
\newcommand{\ee}{\end{equation}}
\newcommand{\bear}{\begin{array}}
\newcommand{\eear}{\end{array}}
\newcommand{\ba}{\begin{eqnarray}}
\newcommand{\ea}{\end{eqnarray}}

\def\a{\alpha}
\def\b{\beta}

\def\d{\delta}

\def\g{\gamma}

\def\l{\lambda}
\def\m{\mu}
\def\n{\nu}

\def\r{\rho}
\def\s{\sigma}

\def\tns{\tensor}

\newcommand{\TeV}{\; \mathrm{TeV}}

\title{ Inflation in Metric-Affine Quadratic Gravity}

\author[a]{Ioannis D.~Gialamas } 
\author[b]{and Kyriakos Tamvakis}

\emailAdd{ioannis.gialamas@kbfi.ee}
\emailAdd{tamvakis@uoi.gr}

\affiliation[a]{Laboratory of High Energy and Computational Physics, 
National Institute of Chemical Physics and Biophysics, R{\"a}vala pst.~10, Tallinn, 10143, Estonia}
\affiliation[b]{Physics Department, University of Ioannina, 45110, Ioannina, Greece}

\abstract{In the general framework of \textit{Metric-Affine} theories of gravity, where the metric and the connection are independent variables, we consider actions quadratic in the Ricci scalar curvature and the Holst invariant (the contraction of the Riemann curvature with the Levi-Civita antisymmetric tensor) coupled non-minimally to a scalar field. We study the profile of the equivalent effective metric theory, featuring an extra dynamical pseudoscalar degree of freedom, and show that it reduces to an effective single-field inflationary model. We analyze in detail the inflationary predictions and find that they fall within the latest observational bounds for a wide range of parameters, allowing for an increase in the tensor-to-scalar ratio. The spectral index can either decrease or increase depending
on the position in parameter space. }

\begin{document}

\maketitle

\vspace{1cm}
\section{Introduction} Cosmological inflation~\cite{AAS, DK, KS, AHG, ADL1, AS, ADL2}, according to which the Universe underwent a phase of quasi-de Sitter expansion, has become central to our present understanding of early Cosmology primarily because it provides a mechanism for the origin of formation of the observed large scale structure~\cite{AAS1, MUK, HAWK, AAS2, AHG2, BST}. A standard way to model inflation is to introduce a scalar field (the {\textit{inflaton}}) that on the one hand provides the vacuum energy driving the expansion and on the other hand through its quantum treatment generates the spatial inhomogeneities that at presently are detected in CMB.

The common framework of all cosmological models is Einstein's General Relativity (GR) with gravity entering in the action through the standard Einstein-Hilbert term and the inflaton coupled minimally to it through the metric. Nevertheless, although gravity is treated classically, the scalar fields are expected to be sensitive to quantum corrections which are likely to generate correction terms as a non-minimal coupling of the scalar fields to the Ricci scalar $\xi\phi^2R$ or quadratic terms like $\gamma R^2$. In fact, the Starobinsky model~\cite{AAS} based on the latter correction has been quite successful in its predictions of inflationary observables~\cite{Akrami:2018odb,BICEP:2021xfz}. Similarly, the non-minimal coupling to the Ricci scalar is central to the {\textit{Higgs inflation}} model~\cite{BS}. The understanding is that these terms, even if they are absent in the classical action, they are bound to be generated and, therefore, they should be included in the effective action employed to model inflationary phenomena.

In the standard (metric) formulation of GR the affine connection is not an independent variable but it is constraint to depend on the metric through the Levi-Civita relation. In the case of the simple Einstein-Hilbert action with minimally coupled matter fields this is entirely equivalent to the so-called \textit{``first order"} or {Palatini} formulation in which the connection is an independent variable~\cite{PALA} (see also~\cite{Sotiriou:2006qn, Borunda:2008kf, Sotiriou:2008rp, DeFelice:2010aj,  Olmo:2011uz, Clifton:2011jh, Capozziello:2011et}). Nevertheless, in the presence of non-minimal corrections, like the ones mentioned above, this no longer true and the two formulations lead to different predictions~\cite{BD}. The Palatini version of the Starobinsky model does not provide any scalar degree of freedom suitable to play the role of the inflaton in contrast to its metric version. Nevertheless, Palatini inflationary models of a scalar field in the presence of an $R^2$ term have been studied with predictions compatible with existing cosmological data~\cite{EERW, AKLT} (see also~\cite{Bombacigno:2018tyw, AKLPT, Tenkanen:2019jiq, Edery:2019txq, Tenkanen:2019wsd, Gialamas:2019nly,   Tenkanen:2020cvw, Lloyd-Stubbs:2020pvx, Antoniadis:2020dfq, Ghilencea:2020piz, Das:2020kff, Tang:2020ovf, Gialamas:2020snr,  Ghilencea:2020rxc, Iosifidis:2018zjj, Iosifidis:2020dck, Bekov:2020dww,   Dimopoulos:2020pas,   Karam:2021sno, Lykkas:2021vax, Saez-ChillonGomez:2021byq,  Gialamas:2021enw,  Antoniadis:2021axu, Annala:2021zdt,  Dioguardi:2021fmr, AlHallak:2022gbv, Dimopoulos:2022tvn, Panda:2022esd,  Dimopoulos:2022rdp,  Durrer:2022emo, Antoniadis:2022cqh, Lahanas:2022mng, Panda:2022can} for various applications.) 

The Palatini formulation of GR coupled to scalar fields falls into the framework of the so-called {\textit{Metric-Affine}} theories of gravity in which the metric and the connection represent independent degrees of freedom~\cite{BALD,HEHL}. The Palatini formulation terminology is used when the matter fields are not coupled to the independent connection. This is usually the case of scalar fields. This generalization of GR to a \textit{Metric-Affine} theory preserves general covariance and is quite natural from the geometric point of view. The difference between the independent connection and the Levi-Civita one turns out to be a tensor $\tns{C}{_\m^\l_\n}$, named the \textit{Distortion}  tensor. In general, solving for the \textit{Distortion}, a \textit{Metric-Affine} theory can be rewritten as an {\textit{``equivalent"}} metric theory which may or may not have the same degrees of freedom as the starting \textit{Metric-Affine} theory, depending on whether the \textit{Distortion} is dynamical or not. In the case that new dynamical degrees of freedom are present in the resulting theory, its phenomenology will be different and possibly interesting~\cite{SALVIO1, SALVIO2}. 

In the present article we considered a \textit{Metric-Affine} theory of gravity based on an action that includes a quadratic  Ricci scalar term ${\cal{R}}^2$ as well as linear and quadratic terms of the Holst invariant $\widetilde{\cal{R}}$, coupled non-minimally to a scalar $\phi$, and studied the inflationary behaviour of the resulting two-scalar field model.  As we have already stated above there is plenty of motivation for the study of quadratic gravitational terms, based on the fact that such terms are bound to be generated by quantum corrections. Therefore, they should be included in the effective theory of gravity to be considered as a framework for the investigation of inflationary phenomena. This is independent of whether the presence of the extra terms will provide additional dynamical degrees of freedom that could either carry out inflation or participate in it through their mixing with a fundamental inflaton. Although that in the present article we focus on the \textit{Metric-Affine} formulation, the motivation applies equally well to the metric formulation as well. Apart from the theoretical interest on the models, their resulting inflationary phenomenology, although in agreement with present observational data, allows for their disapproval or favourable comparison to other models in the light of future more precise data. We have found that the model reduces to an effective single-field model with a potential analogous to the one arising in the Palatini-${\cal{R}}^2$ model with its characteristic inflationary plateau. The inflationary predictions of the model fall within the latest observational bounds for a wide range of parameters.  

In Section~\ref{sec:Framework} we set up the theoretical framework of \textit{Metric-Affine} theories formulated in terms of the \textit{Distortion}  tensor. In Section~\ref{sec:quadratic} we consider \textit{Metric-Affine} theories based on an action that includes quadratic terms of the Ricci scalar ${\cal{R}}^2$ and the Holst invariant $\widetilde{\cal{R}}^2$ and derive the equivalent metric action. In Section~\ref{sec:fundamental} we consider the above quadratic theory coupled non-minimally to a scalar field.  In Section~\ref{sec:inflation} we analyze in detail the inflationary predictions of the above model. Finally,
we summarize and conclude in Section~\ref{sec:conclusions}.

We use natural units $\hbar=c=k_B=1$ as well as taking the reduced Planck mass $M_P=1$.  The metric signature is $(-,+,+,+)$ throughout. 

\section{Framework}
\label{sec:Framework} 
A {\textit{Metric-Affine}} theory is defined in terms of a metric $g_{ \mu\nu}$ and an independent connection $\tns{\widetilde{\Gamma}}{_\m^\r_\n}$. Note that no particular symmetry is assumed for the lower indices of the connection in contrast with the Levi-Civita connection of metric theories where they are symmetric. A curvature tensor is defined as 
\be  \tns{\cal{R}}{_\m_\n^\r_\s}\,\equiv\,\partial_{ \mu}\tns{\widetilde{\Gamma}}{_\n^\r_\s}-\partial_{ \nu}\tns{\widetilde{\Gamma}}{_\m^\r_\s}+\tns{\widetilde{\Gamma}}{_\m^\r_\l}\tns{\widetilde{\Gamma}}{_\n^\l_\s}-\tns{\widetilde{\Gamma}}{_\n^\r_\l}\tns{\widetilde{\Gamma}}{_\m^\l_\s}\,,\ee
The \textit{Distortion} is defined as
\be \tns{C}{_\m^\r_\n}\,\equiv\,\tns{\widetilde{\Gamma}}{_\m^\r_\n}-\tns{\Gamma}{_\m^\r_\n}\,,\ee
where $\tns{\Gamma}{_\m^\r_\n}$ is the Levi-Civita connection
\be \tns{\Gamma}{_\m^\r_\n}\,=\,\frac{1}{2}g^{\rho\sigma}\left(\partial_{ \mu}g_{ \sigma\nu}\,+\,\partial_{ \nu}g_{ \mu\sigma}\,-\partial_{ \sigma}g_{ \mu\nu}\right)\,.\ee
The \textit{Distortion} can be shown to be a tensor. 

The following two scalars, that are linear in the Riemann tensor, can be defined  as
\ba 
\cal{R}&=&\tns{\cal{R}}{_\m_\n^\r_\s}\d_{\rho}^{ \mu}g^{\nu\sigma}\,=\,\tns{\cal{R}}{_\m_\n^\m^\n}  \,,
\\ \widetilde{\cal{R}}&=&(-g)^{-1/2}\tns{\epsilon}{^\m^\n_\r^\s} \tns{\cal{R}}{_\m_\n^\r_\s}=(-g)^{-1/2}\tns{\epsilon}{^\m^\n^\r^\s}\tns{\cal{R}}{_\m_\n_\r_\s}\,,
\ea
 where $g$ is the determinant of the metric and $\epsilon^{\m\n\r\s}$ is the totally antisymmetric symbol with $\epsilon^{0123}=1$. The first scalar is the usual Ricci scalar and the second one is the so-called Holst invariant. Note that the pseudoscalar $\widetilde{\cal{R}}$~\cite{Hojman:1980kv,Holst:1995pc} vanishes identically in a metric theory due to the symmetry in the lower indices of the Levi-Civita connection. Both these scalars can be written in terms of the standard (metric) Ricci scalar $R[g]$ and the \textit{Distortion} $C$ as\footnote{The indices of the \textit{Distortion} tensor are raised and lowered in the usual way, \textit{i.e} $\tns{C}{_\a^\b^\g}=\tns{C}{_\a^\b_\d}g^{\gamma\delta}$ and $\tns{C}{_\a_\b_\g}\,=\,g_{ \beta\delta}\tns{C}{_\a^\d_\g}$.} 
\ba 
\label{eq:ricci_C}
{\cal{R}}&=&R+D_{ \mu}\tns{C}{_\n^\m^\n}-D_{ \nu}\tns{C}{_\m^\m^\n}+\tns{C}{_\m^\m_\l}\tns{C}{_\n^\l^\n}-\tns{C}{_\n^\m_\l}\tns{C}{_\m^\l^\n}\,,
\\
\widetilde{\cal{R}}&=&2(-g)^{-1/2}\epsilon^{ \mu\nu\rho\sigma}\left(D_{ \mu}\tns{C}{_\n_\r_\s}+\tns{C}{_\m_\r_\l}\tns{C}{_\n^\l_\s}\right)\,,
\label{eq:Holst_C}
\ea
where the covariant differentiation is in terms of the Levi-Civita connection. 

The simplest \textit{Metric-Affine} action is the analogue of the Einstein-Hilbert action
\ba
{\cal{S}}&=&\frac{1}{2}\int {\rm d}^4x\,\sqrt{-g}{\cal{R}} \nonumber
\\ &=& \frac{1}{2}\int{\rm d}^4x\,\sqrt{-g}\left\{R+D_{ \mu}\tns{C}{_\n^\m^\n}-D_{ \nu}\tns{C}{_\m^\m^\n}+\tns{C}{_\m^\m_\l}\tns{C}{_\n^\l^\n}-\tns{C}{_\n^\m_\l}\tns{C}{_\m^\l^\n}\right\}\,.{\label{MAEH}}
\ea
Variation with respect to the \textit{Distortion} gives the equation of motion  
\be\frac{\delta {\cal{S}}}{\delta C}\,=\,0\,\,\,\Longrightarrow\,\,\delta_{ \beta}^{ \alpha}\tns{C}{_\nu_\gamma^\nu}+\delta_{ \gamma}^{ \alpha}\tns{C}{_\nu^\nu_\beta}-\tns{C}{_\beta_\gamma^\alpha}-\tns{C}{_\gamma^\alpha_\beta}=0\,,
\ee
which has the general solution $C_{\mu\nu\rho}=U_{\mu}\,g_{\nu\rho}\,$
in terms of the arbitrary vector $U_{ \mu}$. Substituting this solution into Eq.~\eqref{MAEH} the $C-$dependent terms vanish. Therefore, the \textit{Metric-Affine} theory ~\eqref{MAEH} is entirely equivalent to standard metric GR for any $U_{ \mu}$. Nevertheless, this is not the case for quadratic actions.

Closing this section we mention that the terms \textit{Metric-Affine} and \textit{Palatini formulation} are equivalent as far as they are referring to gravity theory with an independent connection, while torsion and non-metricity being an immediate consequence of that.

\section{Quadratic \textit{Metric-Affine} theories}
\label{sec:quadratic}
Consider the following generalization of the Starobinsky model
\be {\cal{S}}\,=\,\int{\rm d}^4x\,\sqrt{-g}\left\{\frac{\alpha}{2}{\cal{R}}\,+\,\frac{\beta}{2} \widetilde{\cal{R}}\,+\,\frac{\gamma}{4}{\cal{R}}^2\,+\,\frac{\delta}{4}\widetilde{\cal{R}}^2\right\}\,.{\label{ACT-00}}\ee
 Without loss of generality we may set $\alpha=1$,  \textit{i.e.} we identify the parameter $\alpha$ with the reduced Planck mass and rewrite the action in the form

\be {\cal{S}}\,=\,\int{\rm d}^4x\sqrt{-g}\left\{\frac{1}{2}(1+\gamma \chi){\cal{R}}+\frac{1}{2}(\beta+\delta\zeta)\widetilde{\cal{R}}-\frac{\gamma}{4}\chi^2-\frac{\delta}{4}\zeta^2\,\right\}\,,{\label{ACT-0}}\ee
introducing the two auxiliary scalars $\chi$ and $\zeta$. As we will see a dynamical scalar field can be generated in this case. Using ~\eqref{eq:ricci_C} and~\eqref{eq:Holst_C} of the previous section we can rewrite the action in terms of the \textit{Distortion} $C$ as
\ba
\cal{S}&=&\int{\rm d}^4x\sqrt{-g}\left\{\frac{1}{2}(1+\gamma \chi){{R}} +\frac{1}{2}(1+\gamma\chi)\left(D_{ \mu}\tns{C}{_\n^\m^\n}-D_{ \nu}\tns{C}{_\m^\m^\n}+\tns{C}{_\m^\m_\l}\tns{C}{_\n^\l^\n}-\tns{C}{_\n^\m_\l}\tns{C}{_\m^\l^\n}\right)\right.\nonumber
\\ && \left.+(\beta+\delta\zeta)(-g)^{-1/2}\epsilon^{ \mu\nu\rho\sigma}\left(D_{ \mu}\tns{C}{_\n_\r_\s}+\tns{C}{_\m_\r_\l}\tns{C}{_\n^\l_\s}\right) -\frac{\gamma}{4}\chi^2-\frac{\delta}{4}\zeta^2\right\}\,.
\label{ACT-1}
\ea
Substituting the covariant derivatives of the \textit{Distortion} this is equivalent to
\ba
\cal{S} &=& \int{\rm d}^4x\sqrt{-g}\left\{\frac{1}{2}(1+\gamma \chi){{R}}-\frac{\gamma}{4}\chi^2-\frac{\delta}{4}\zeta^2-\frac{1}{\sqrt{-g}}\partial_{ \mu}\left(\frac{1}{2}\sqrt{-g}(1+\gamma\chi)\right)\tns{C}{_\n^\m^\n}\right. \nonumber
\\ && +\frac{1}{\sqrt{-g}}\partial_{ \nu}\left(\frac{1}{2}\sqrt{-g}(1+\gamma\chi)\right)\tns{C}{_\m^\m^\n}+\frac{1}{2}(1+\gamma\chi)\left(\tns{C}{_\m^\m_\l}\tns{C}{_\n^\l^\n}-\tns{C}{_\n^\m_\l}\tns{C}{_\m^\l^\n}\right. \nonumber
\\ && \left.+\tns{\Gamma}{_\m^\m_\r}\tns{C}{_\n^\r^\n}+\tns{\Gamma}{_\m^\n_\r}\tns{C}{_\n^\m^\r} -\tns{\Gamma}{_\n^\m_\r}\tns{C}{_\m^\r^\n}-\tns{\Gamma}{_\n^\n_\r}\tns{C}{_\m^\m^\r} \right)-\frac{1}{\sqrt{-g}}\partial_{ \mu}(\beta+\delta\zeta)\epsilon^{ \mu\nu\rho\sigma}C_{ \nu\rho\sigma}\nonumber
\\ && +(\beta+\delta\zeta)(-g)^{-1/2}\epsilon^{ \mu\nu\rho\sigma}\tns{C}{_\m_\r_\l}\tns{C}{_\n^\l_\s}\bigg\}\,.
\label{ACT-2}
\ea
Variation with respect to $\tns{C}{_\a^\b^\g}$ gives
\be 
\frac{\Omega^2}{2}\left(\delta_{ \beta}^{ \alpha}\tns{C}{_\n_\g^\n}+\delta_{ \gamma}^{ \alpha}\tns{C}{_\n^\n_\b}-\tns{C}{_\b_\g^\a}-\tns{C}{_\g^\a_\b}\right)
+\frac{\overline{\Omega}^2}{\sqrt{-g}}\left(-\tns{\epsilon}{^\m^\a^\s_\b}\tns{C}{_\m_\g_\s}-\tns{\epsilon}{^\m^\a^\s_\g}\tns{C}{_\m_\s_\b}\right)\,=\,\tns{J}{_\b^\a_\g}\,,
\label{CEQ-1}
\ee
with
\be
\tns{J}{_\b^\a_\g}\,=\,\frac{1}{2}\partial_{ \beta}\Omega^2\,\delta_{ \gamma}^{ \alpha}\,-\frac{1}{2}\partial_{ \gamma}\Omega^2\,\delta_{ \beta}^{ \alpha}+\frac{1}{\sqrt{-g}}\partial_{ \mu}\overline{\Omega}^2 \tns{\epsilon}{^\m^\a_\b_\g}.
\ee 
The conformal factors $\Omega^2, \overline{\Omega}^2 $ are given by $ \Omega^2\,\equiv 1+\gamma\chi$ and $\overline{\Omega}^2\,\equiv\,\beta+\delta\zeta$.
Note also that the tensor $\tns{J}{_\b^\a_\g}$ is antisymmetric in its lower indices, \textit{i.e.} $\tns{J}{_\b^\a_\g}\,=\,-\tns{J}{_\g^\a_\b}$.
Thus, we proceed by considering a trial solution for $C_{ \mu\nu\rho}$ antisymmetric in the last two indices, namely
\be C_{ \mu\nu\rho}\,=\,g_{ \mu\nu}\partial_{ \rho}X-g_{\mu\rho}\partial_{ \nu}X+\epsilon_{ \mu\nu\rho\sigma}\partial^{ \sigma}Y\,.\ee
Introducing this ansatz into {\eqref{CEQ-1}} we obtain
\be \partial X\,=\,\frac{1}{2\Delta}\left(\Omega^2\partial\Omega^2+4\overline{\Omega}^2\partial\overline{\Omega}^2\right)\,,\,\,\,\,\,\,\,\partial Y\,=\,\frac{1}{\sqrt{-g}\Delta}\left(\Omega^2\partial\overline{\Omega}^2-\overline{\Omega}^2\partial\Omega^2\right)\,,\ee
with $\Delta\,=\,\Omega^4+4\overline{\Omega}^4$.

Substituting the solution for the \textit{Distortion} into the action ({\ref{ACT-1}}) we obtain

\be {\cal{S}}\,=\,\int{\rm d}^4x\,\sqrt{-g}\left\{\,\frac{1}{2}\Omega^2 R\,+\,\frac{3}{4}\frac{(\nabla\Omega^2)^2}{\Omega^2}\,-\frac{3}{\Omega^2\Delta}\left(\Omega^2\nabla\overline{\Omega}^2-\overline{\Omega}^2\nabla\Omega^2\right)^2\,-\frac{\gamma}{4}\chi^2-\frac{\delta}{4}\zeta^2\,\right\}\,.\ee

 The final step is the passage to the Einstein-frame applying a Weyl rescaling of the form  $g_{ \mu\nu}\,=\,\bar{g}_{ \mu\nu}/\Omega^2$. Note that the standard (metric) Ricci scalar is rescaled as $R=\Omega^2 \bar{R} -6\Omega^3 \Box \Omega^{-1}$, in contrast to the \textit{Metric-Affine} one which follows the rule $\mathcal{R}=\Omega^2 \bar{\mathcal{R}}$, since the curvature $\tns{\cal{R}}{_\m_\n^\r_\s}$ is metric-independent. 
Under this Weyl rescaling 
we obtain the Einstein-frame action
\be {\cal{S}}\,=\,\int{\rm d}^4x\,\sqrt{-\bar{g}}\left\{\frac{1}{2}\bar{R}[\bar{g}]\,-\frac{3}{\Omega^4\Delta}\left(\Omega^2\bar{\nabla}\overline{\Omega}^2-\overline{\Omega}^2\bar{\nabla}\Omega^2\right)^2\,-\frac{1}{4}\frac{\gamma\chi^2+\delta\zeta^2}{\Omega^4}\right\}\,.\ee
Introducing the field
\be \sigma\,\equiv\,\frac{\overline{\Omega}^2}{2\Omega^2}\,.\ee
we obtain the scalar field Lagrangian 
\be {\cal{L}}\,=\,-\frac{12(\bar{\nabla}\sigma)^2}{(1+16\sigma^2)}-\frac{1}{4\gamma}\left(1-\Omega^{-2}\right)^2-\frac{1}{4\delta}\left(2\sigma-\beta\Omega^{-2}\right)^2\,.\ee 
Solving with respect to the non-dynamical $\Omega^2$ we get
\be \frac{\delta{\cal{L}}}{\delta\Omega^2}=0\,\,\,\,\Longrightarrow\,\,\,\Omega^{-2}=\frac{\delta+2\beta\gamma\sigma}{\delta+\beta^2\gamma}\,,\ee
which leads to 
\be {\cal{L}}\,=\,-\frac{12(\bar{\nabla}\sigma)^2}{(1+16\sigma^2)}-\frac{1}{4}\frac{(2\sigma-\beta)^2}{(\delta+\beta^2\gamma)}\,.\ee
Thus, the model based on the action~\eqref{ACT-00}, in addition to the standard graviton, predicts a pseudoscalar gravitational degree of freedom associated with the invariant $\widetilde{\cal{R}}$. 
The model can be expressed in terms of a canonical field $s$, defined by
\be \sigma=\frac{1}{4}\sinh(\sqrt{2/3}\,s)\,.\ee 
The potential becomes
\be V(s)\,=\,\frac{1}{16}\frac{\left(\sinh(\sqrt{2/3} s)-2\beta\right)^2}{(\delta+\beta^2\gamma)}\,.\ee
This model and its inflationary behaviour have been considered in~\cite{SALVIO1,SALVIO2}.
If we were to take the limit $\beta\rightarrow 0$, while $\delta\neq 0$, we would not be able to obtain an acceptable inflationary behaviour as in the case of a non-zero $\beta$.  Thus, the parameter\footnote{The parameter $1/(2\beta)$ is called Barbero-Immirzi parameter~\cite{Immirzi:1996dr,Immirzi:1996di}.} $\beta$, associated with the linear Holst term, is crucial  in obtaining a suitable inflationary phenomenology, since it enhances the flatness of the potential in a certain region. It is evident that the parameters $\gamma$ and $\delta$, associated with the presence of the ${\cal{R}}^2$ and $\widetilde{\cal{R}}^2$ term, have a secondary role in the inflationary behaviour of the model, which is controled by $\beta$, while $\gamma$ and $\delta$ affect only the scale of inflation. Even without one of them, the inflationary model would still be viable, since even a single term is sufficient to achieve the desired inflationary scale. Note however, that the above pseudoscalar mode would not be present if the action were to contain only the linear Holst term. At least one of the ${\cal{R}}^2$ or $\widetilde{\cal{R}}^2$ {{has}} to be included in order to generate a dynamical scalar degree of freedom~\cite{SALVIO1,SALVIO2}. In the absence of $\widetilde{\cal{R}}${{, i.e. $\beta=0$}}, a quadratic Holst term $\widetilde{\cal{R}}^2$ can generate a dynamical scalar as well, but in this case no viable inflation arises. Note also that all this is in contrast to the \textit{Metric-Affine} version of the Starobinsky model (just an ${\cal{R}}^2$ term present), where no dynamical scalar degree of freedom arises. In Table~\ref{TABLE:CASES} we list the possible combinations of curvature scalars up to quadratic terms in the framework of \textit{Metric-Affine} gravity and indicate the presence or absence of the new dynamical scalar in the Metric equivalent theory as well as the inflationary viability of the corresponding model. 
\begin{table}[t!]
\caption{List of different combinations of the curvature scalars in \textit{Metric-Affine}  gravity (upper row) and the equivalent Metric theory (lower row). VI (\textbf{N}VI) stands for viable (not viable) inflation and $\sigma$ is the dynamical scalar.}
\small
\makebox[1 \textwidth][c]{       
\resizebox{1.015 \textwidth}{!}{   
\begin{tabular}{m{3.35em}| c c c c c c c c c}
\hline \hline
\textit{Metric-Affine} Gravity  & $\mathcal{R}$ & $\mathcal{R}+\widetilde{\mathcal{R}}$ & $\mathcal{R}+\mathcal{R}^2$ & $\mathcal{R} + \widetilde{\mathcal{R}}^2$ & $\mathcal{R}+ \widetilde{\mathcal{R}} +\mathcal{R}^2$ & $\mathcal{R} + \widetilde{\mathcal{R}}+ \widetilde{\mathcal{R}}^2$ & $\mathcal{R} +\widetilde{\mathcal{R}}^2 +\mathcal{R}^2$ & $\mathcal{R} +\widetilde{\mathcal{R}} +\widetilde{\mathcal{R}}^2 +\mathcal{R}^2$
\\
\hline
\textit{Metric} Gravity & GR & GR & GR  & $\sigma$ with \textbf{N}VI & $\sigma$ with VI & $\sigma$ with VI & $\sigma$ with \textbf{N}VI & $\sigma$ with VI \\
\hline \hline
\end{tabular}
}}
\label{TABLE:CASES}
\end{table}

\section{Coupling to a fundamental scalar}
\label{sec:fundamental}
Assuming that we have a model of a scalar field $\phi$ coupled to gravity in the framework of \textit{Metric-Affine} theory, as it was mentioned in the introduction non-minimal terms as the non-minimal coupling to the curvature scalars or quadratic scalar curvature terms are bound to be present in the effective theory. Therefore, we start by considering
\be
\label{eq:action1n}
S=\int {\rm d}^4 x \sqrt{-g}\left\{\frac{1}{2}f(\phi)\mathcal{R} +\frac{1}{2}g(\phi)\widetilde{\cal{R}} +\frac{\gamma}{4} \mathcal{R}^2+\frac{\delta}{4} \widetilde{R}^2 + \mathcal{L}_\phi\right\}\,,
\ee
with
\be
\label{eq:matter_Lagn}
\mathcal{L}_\phi = -\frac{1}{2}g^{\m\n}\partial_\m \phi \partial_\n \phi -V(\phi)\,.
\ee
This can be rewritten in terms of the  auxiliary scalars $\chi$ and $\zeta$ as
\be
\label{eq:action2n}
S=\int {\rm d}^4 x \sqrt{-g}\left\{\frac{1}{2}(\gamma \chi+f(\phi) ) \mathcal{R}+\frac{1}{2} \left(\d\zeta +g(\phi)\right)\widetilde{\mathcal{R}}-\frac{\gamma}{4} \chi^2-\frac{\delta}{4} \zeta^2 +\mathcal{L}_\phi\right\}\,.
\ee
Defining $\Omega^2=\g\chi + f(\phi)$, $\overline{\Omega}^2 = \d \zeta +g(\phi)$ and following the previous section we obtain
\be
\label{eq:action5n}
\mathcal{S}=\int {\rm d}^4 x \sqrt{-g} \left\{\frac{R}{2} -
\frac{3}{\Omega^4 \Delta}\left(\Omega^2 \partial \overline{\Omega}^2 -\overline{\Omega}^2 \partial \Omega^2 \right)^2 -\frac{\gamma}{4\Omega^4} \chi^2-\frac{\delta}{4\Omega^4} \zeta^2 -\frac{1}{2}\frac{(\nabla\phi)^2}{\Omega^2}-\frac{V(\phi)}{\Omega^4}\right\}\,,
\ee
where now we have to substitute $\d\zeta^2 = \left(\overline{\Omega}^2 -g(\phi) \right)^2/\d$ and $\g\chi^2 = \left(\Omega^2 -f(\phi) \right)^2/\g$. Introducing the scalar field $\sigma=\frac{\overline{\Omega}^2}{2\Omega^2}$ we obtain 
\ba
\label{eq:action6n}
\mathcal{S}&=&\int {\rm d}^4 x \sqrt{-g} \bigg\{\frac{R}{2} -\frac{24}{1+16\s^2}\frac{(\nabla\s)^2}{2}-\frac{(\nabla\phi)^2}{2\Omega^2}-\frac{\s^2}{\d} \nonumber
\\ && -\frac {(f(\phi) - \Omega^2)^2}{4\g \Omega^4}-\frac{g(\phi)}{4\delta\Omega^4}(g(\phi)-4\sigma\Omega^2)-\frac{V(\phi)}{\Omega^4} \bigg\}\,.
\ea
Varying the action with respect to $\Omega^2$ we obtain from the solution of $\d\mathcal{S}/\d\Omega^2=0$, that 
\be
\label{eq:omega2}
\Omega^2=\frac{4\gamma V(\phi)+f^2(\phi)+\gamma g^2(\phi)/\delta}{f(\phi)+2\gamma\sigma g(\phi)/\delta-\g(\nabla\phi)^2}\,.
\ee
Substituting this back to the action~\eqref{eq:action6n} we get
\be
\label{eq:action8n}
\mathcal{S}=\int {\rm d}^4 x \sqrt{-g} \bigg\{\frac{R}{2} - K_\phi(\phi,\s)\frac{(\nabla\phi)^2}{2} +L_\phi(\phi)\frac{(\nabla\phi)^4}{4} -K_\s(\s) \frac{(\nabla\s)^2}{2}-V_{\rm eff}(\phi,\s)\bigg\}\,,
\ee
with
\ba
K_\phi(\phi,\s) &=& \frac{f(\phi)+2\g\s g(\phi)/\d}{\gamma g(\phi)^2/\delta+f^2(\phi) +4\g V(\phi)}\,,\nonumber
\\
L_\phi(\phi) &=&  \frac{\g}{\gamma g(\phi)^2/\delta+f^2(\phi)+4\g V(\phi)}\,, \nonumber
\\
K_\s(\s) &=& \frac{24}{1+16\s^2}\,, \nonumber
\\
 V_{\rm eff}(\phi,\s) &=& \frac{V(\phi)}{f^2(\phi)+4\gamma V(\phi)}+\frac{1}{\delta}\left(\frac{f^2(\phi)+4\gamma V(\phi)}{\gamma g^2(\phi)/\delta+f^2(\phi)+4\gamma V(\phi)}\right)\left(\sigma-\sigma_0(\phi)\right)^2\,.
 \label{eq:3dpot}
\ea
The potential $V_{eff}(\phi,\sigma)$ is positive-definite\footnote{The potential $V(\phi)$ and the parameters $\gamma$ and $\delta$ are taken to be positive.} and gets minimized along the line
\be \sigma_0(\phi)\,=\,\frac{g(\phi)f(\phi)/2}{f^2(\phi)+4\gamma V(\phi)}\,.
\label{eq:sigma_0}
\ee
Note that along this line the potential coincides with that obtained in the $\delta=g(\phi)=0$ case~\cite{EERW, AKLT, AKLPT}.
A canonical $\sigma$-field can readily be defined as
\be \sigma_c\,=\,\sqrt{24}\int\,\frac{ {\rm d}\sigma}{\sqrt{1+16\sigma^2}}\,\Longrightarrow\,\sigma\,=\,\frac{1}{4}\sinh\left(\sqrt{\frac{2}{3}}\sigma_c\right)\,.
\label{eq:sigma_can}
\ee
The $\phi$-dependent mass of $\sigma_c$ along the minimum line is
\ba
M_{\sigma_c}^2(\phi)&\equiv&\left.\frac{\partial^2V_{eff}}{\partial\sigma_c^2}\right|_{\sigma_0}=\left.\left(\frac{\partial\sigma}{\partial\sigma_c}\right)^2\frac{\partial^2V_{eff}}{\partial\sigma^2}\right|_{\sigma_0}\,+\,{\cancel{\left.\frac{\partial^2\sigma}{\partial\sigma_c^2}\frac{\partial V_{eff}}{\partial\sigma}\right|_{\sigma_0}}} \Rightarrow \nonumber
\\ M_{\sigma_c}^2(\phi)&=&\frac{\left(f^2(\phi)+4\gamma V(\phi)\right)^2+4g^2(\phi)f^2(\phi)}{12\delta\left(f^2(\phi)+4\gamma V(\phi)\right)\left(\gamma g^2(\phi)/\delta+f^2(\phi)+4\gamma V(\phi)\right)}\,>\,0\,.
\ea
Assuming a quartic potential and a quadratic $f(\phi)$ function, this mass tends to a constant for $\phi\rightarrow\,\infty$, namely
\be \,M_{\sigma_c}^2(\phi\rightarrow \infty)\,\approx\,\left\{\bear{cc}
\frac{1}{3\gamma}\frac{f^2(\phi)}{f^2(\phi)+4\gamma V(\phi)}\,,&\,g^2(\phi)\gg\phi^4\\
\,&\,\\
\frac{1}{12\delta}\,,&\,g^2(\phi)\ll\phi^4\,.
\eear\right.\ee

Along the minimum line $\sigma_0(\phi)$ the contribution of $\sigma$ to the potential is removed but it still has contribution to the kinetic terms. Substituting $\sigma_0(\phi)$ into the $\phi$-kinetic term, it reduces to the one obtained in the $\delta=g(\phi)=0$ case, namely
\be -\frac{1}{2}\left(\frac{f(\phi)}{f^2(\phi)+4\gamma V(\phi)}\right)(\nabla\phi)^2\,.{\label{PHI-KIN}}\ee
Substituting $\sigma(\phi)$ at the minimum in the $\sigma$ kinetic term we obtain the correction
\ba
-K_\s(\s_0(\phi)) \frac{(\nabla\s_0(\phi))^2}{2}&=&-\frac{1}{2}\left(\frac{12}{1+\frac{4g(\phi)^2f^2(\phi)}{[f^2(\phi)+4\gamma V(\phi)]^2}}\right)(\nabla\phi)^2\left(\frac{g'(\phi)f(\phi)+g(\phi)f'(\phi)}{f^2(\phi)+4\gamma V(\phi)}\right.\nonumber
\\ && \left.-\frac{g(\phi)f(\phi)}{[f^2(\phi)+4\gamma V(\phi)]^2}\left(2f'(\phi)f(\phi)+4\gamma V'(\phi)\right)\right)^2\,.
\label{SIGMA-KIN}
\ea
Thus, the overall $\phi$-kinetic term will be the sum of~\eqref{PHI-KIN} and~\eqref{SIGMA-KIN}.

\begin{figure}[t!]  
\begin{center}
\includegraphics[scale=0.8]{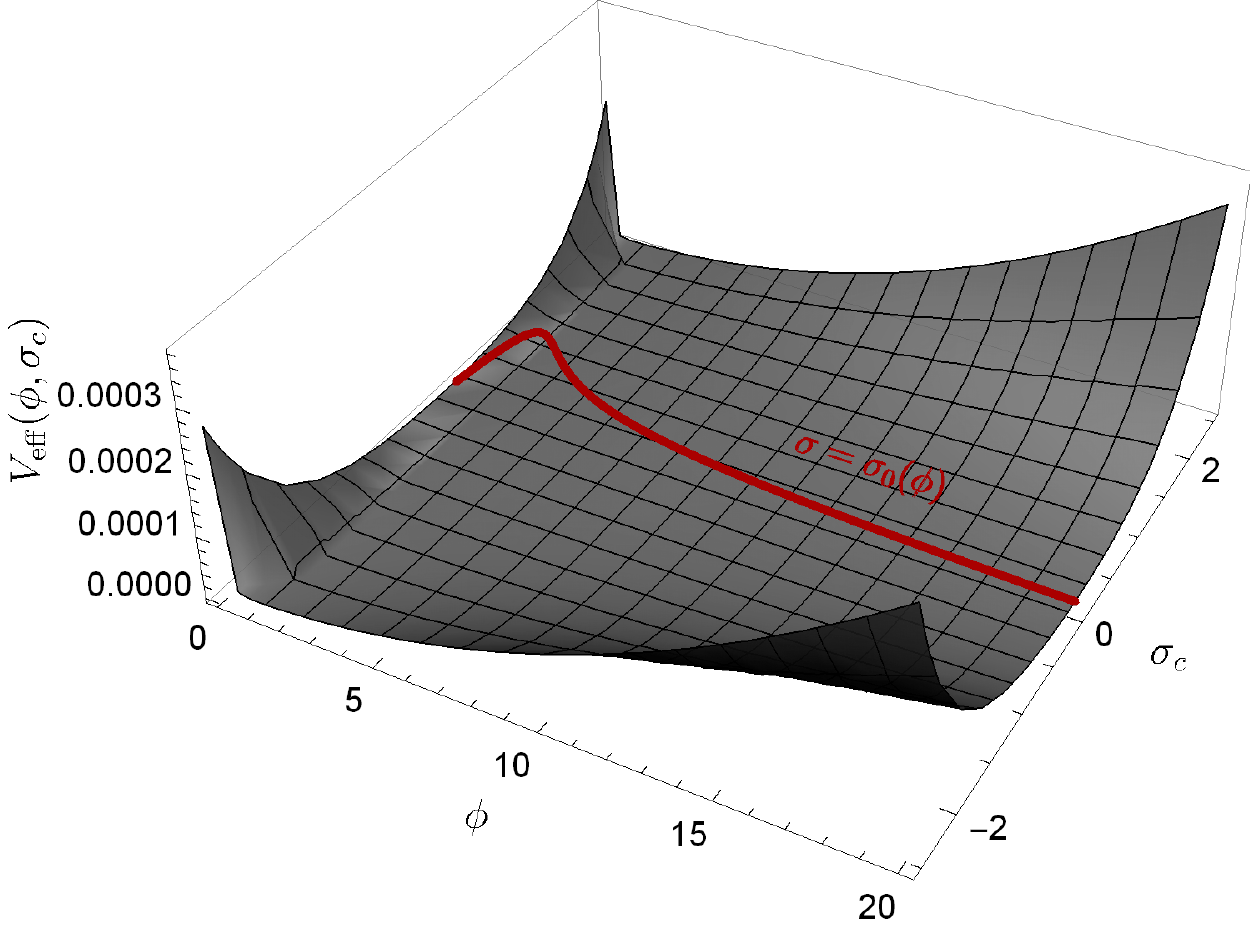}
\caption{The two-field effective potential given by Eq.~\eqref{eq:3dpot}. The parameters used are $\g=10^6$,  $\d=\xi=\bar{\xi}=1$ and $\bar{\xi}' = 0$, while the quartic coupling $\lambda$ is fixed by the observed value of the amplitude of the scalar power spectrum, $A_s^\star=2.1 \times 10^{-9}$, at the pivot scale $k_\star=0.05\, {\rm Mpc^{-1}}$ in the $\s_0(\phi)$ direction (red line).}
\label{fig:3dplot}
\end{center}
\end{figure}

At this point we may proceed to specify the coupling functions\footnote{See also~\cite{Rigouzzo:2022yan} for a general discussion on non-minimal couplings to the torsion and non-metricity tensors.} of $\phi$ with ${\cal{R}}$ and $\widetilde{\cal{R}}$, namely $f(\phi)$ and $g(\phi)$. The standard non-minimal coupling to the Ricci scalar adopted is a quadratic function of $\phi$ that reduces to a constant in the absence of $\phi$. In our case this would correspond to taking
\be f(\phi) = 1+\xi\phi^2\,. \label{eq:f_func}\ee
Next, for the case of the $\widetilde{\cal{R}}$ term, since no such term is present in the minimal gravitational action and is expected to arise as a quantum correction attributed to $\phi$, it is reasonable to assume that $g(\phi)$ should vanish in the absence of it. In addition, since $\widetilde{\cal{R}}$ is parity-odd, it would also be reasonable to assume that $g(\phi)$ is an odd function of $\phi$ (see~\cite{Langvik:2020nrs,Shaposhnikov:2020gts} for quadratic non-minimal couplings to the Holst invariant). Therefore, we shall adopt the following ansatz, namely\footnote{The parameters $\bar{\xi},\,\bar{\xi}'$, as well as the above parameter $\xi$ are assumed to be positive.}
\be g(\phi)=\bar{\xi}\phi+\bar{\xi}'\phi^3\,.
\label{eq:g_func}
\ee 

In figure~\ref{fig:3dplot} we display the two-field effective potential of Eq.~\eqref{eq:3dpot} (for the canonical normalized pseudoscalar $\sigma_c$) using the coupling functions~\eqref{eq:f_func}-\eqref{eq:g_func} and the quartic potential $V(\phi)=\frac{\lambda}{4}\phi^4$. For illustrative purposes we have chosen a linear dependence on $\phi$ for the function $g(\phi)$, \textit{i.e.} $\bar{\xi}'=0$. The values of the rest of the parameters are\footnote{The parameter $\delta$ will be set equal to one from now on without any further reference to it. It turns out that, along the minimum line, where we study inflation, it appears only on the higher order kinetic term which is negligible anyway.} $\g=10^6$ and $\d=\xi=\bar{\xi}=1$. The parameter $\lambda$ is fixed by the observed value of the amplitude of the scalar power spectrum, $A_s^\star=2.1 \times 10^{-9}$, at the pivot scale $k_\star=0.05\, {\rm Mpc^{-1}}$. The red curve corresponds to the minimum direction of Eq.~\eqref{eq:sigma_0}. As it is evident from this figure, in the $\s_c$ direction the potential grows exponentially as $\s_c$ gets larger, hence the pseudoscalar is doomed to fall into the valley. This will be verified in the framework of the numerical treatment of the next section. 
\begin{figure}[t!]  
\begin{center}
\includegraphics[scale=0.8]{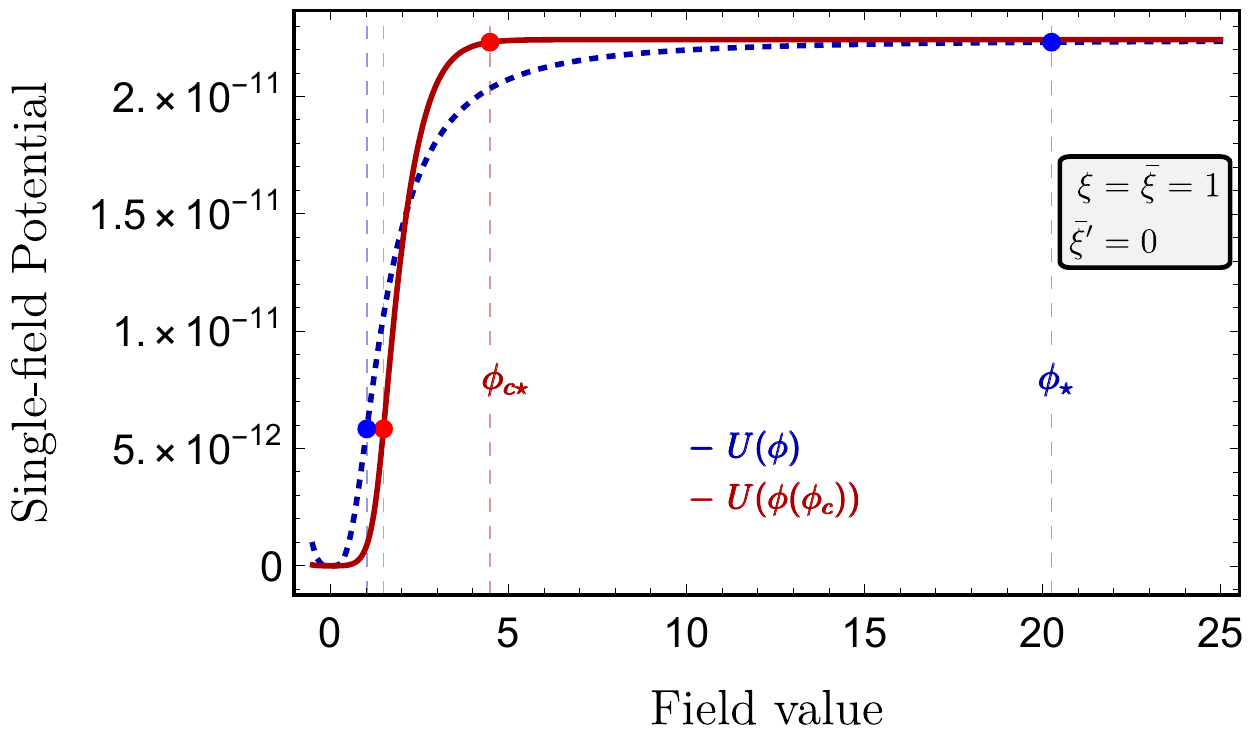}
\caption{The single-field effective potential in the $\s_0(\phi)$ direction. The parameters used coincide with those in figure~\ref{fig:3dplot}. The blue dotted line corresponds to the potential $U(\phi)$ plotted against the non-canonical field $\phi$, while the red one to $U(\phi(\phi_ c))$ plotted in terms of the canonical field $\phi_c$. The dots from right to left give the field values at the pivot scale $k_\star=0.05\, {\rm Mpc^{-1}}$ and at the end of inflation. }
\label{fig:1dplot}
\end{center}
\end{figure}

It is interesting to have a look at the resulting effective Lagrangian along the $\sigma_0(\phi)$ line in the large field limit. Taking the potential to be a standard renormalizable quartic potential $V(\phi)=\frac{\lambda}{4}\phi^4$, we obtain 
\be {\cal{L}}(\phi\rightarrow\infty)\,\approx\,-\frac{1}{2}\frac{(\nabla\phi)^2}{\phi^2}\left(\frac{3}{2} + \frac{1}{\xi+\frac{\gamma\lambda}{\xi}}\right)\,-\frac{\lambda}{4(\xi^2+\gamma\lambda)}\,.{\label{EFFLANG}}\ee 
The potential shows the familiar Palatini-plateau, while the kinetic term is expressed in terms of an effective $\xi$-parameter
\be \xi_{\rm eff}^{-1}\,=\,\frac{3}{2}+\frac{1}{\xi+\frac{\lambda\gamma}{\xi}}\,.\ee
Note that $\xi_{\rm eff}$ is less than $1$. 
We have not displayed in ({\ref{EFFLANG}}) the quartic kinetic term, anticipating that it is not expected to play any role in inflationary considerations, something that will be supported by our subsequent numerical analysis.

In figure~\ref{fig:1dplot} we have plotted the single-field effective potential results when the pseudoscalar $\sigma$ moves along the minimum line $\sigma_0(\phi)$, \textit{i.e.} $U(\phi)=V_{\rm eff}(\phi,\s_0(\phi))$. The values of the parameters have been taken to be the same with those in figure~\ref{fig:3dplot}. Since the inversion from the non-canonical field $\phi$ to the canonical one $\phi_c$ cannot be achieved analytically, we plot $U(\phi)$ (blue dotted line) and $U(\phi(\phi_c))$ (red line) doing the inversion to the canonical field numerically.

\section{Inflation}
In this section we analyze the inflationary behaviour of the model. We aim at restricting the parametric space, comparing its inflationary predictions to the corresponding latest observational bounds as set by  the latest
combination of Planck, BICEP/Keck and BAO data~\cite{Akrami:2018odb,BICEP:2021xfz}.
\label{sec:inflation}
\subsection{Single-field inflation}
Our system of two fields is described by the action ({\ref{eq:action8n}}), where the functions\footnote{For simplicity we have omitted the arguments of these functions.} $K_{ \phi},\,L_{\phi}$ and $V_{eff}$ are given by ({\ref{eq:3dpot}}). Starting from this action the equations of motion arising from it in an FRW background are\footnote{Notice here that we use the canonical normalized pseudoscalar, given by Eq.~\eqref{eq:sigma_can}.}
\ba
&&(K_{ \phi}+3L_{ \phi}\dot{\phi}^2)\ddot{\phi}+3H(K_{ \phi}+L_{ \phi}\dot{\phi}^2)\dot{\phi} + \dot{\phi} \dot{\sigma}_{c} \frac{\partial K_{ \phi}}{\partial\sigma_{c}} +\left(\frac{1}{2}\frac{\partial K_{ \phi}}{\partial\phi}+\frac{3}{4}\frac{\partial L_{ \phi}}{\partial\phi}\dot{\phi}^2\right)\dot{\phi}^2+\frac{\partial V_{eff}}{\partial\phi}=0\,, \nonumber
\\&&\ddot{\sigma}_{c}+3H\dot{\sigma}_{c}-\frac{1}{2}\frac{\partial K_{ \phi}}{\partial\sigma_{c}}\dot{\phi}^2+\frac{\partial V_{eff}}{\partial\sigma_{c}}=0\,, \nonumber
\\ && H^2=\frac{1}{3}\rho\,, \nonumber
\\ && \dot{H}\,=\,-\frac{1}{2}\left(\rho\,+\,p\right)\,,
\ea
where the total energy density and pressure are
\be
\rho =\frac{1}{2}K_{\phi}\dot{\phi}^2+\frac{3}{4}L_{\phi}\dot{\phi}^4+\frac{1}{2}\dot{\sigma}_{c}^2+V_{eff} \quad \text{and} \quad p=\frac{1}{2}K_{\phi}\dot{\phi}^2+\frac{1}{4}L_{\phi}\dot{\phi}^4+\frac{1}{2}\dot{\sigma}_{c}^2-V_{eff}
\,.
\ee

\begin{figure}[t!]  
\begin{center}
\includegraphics[scale=0.541]{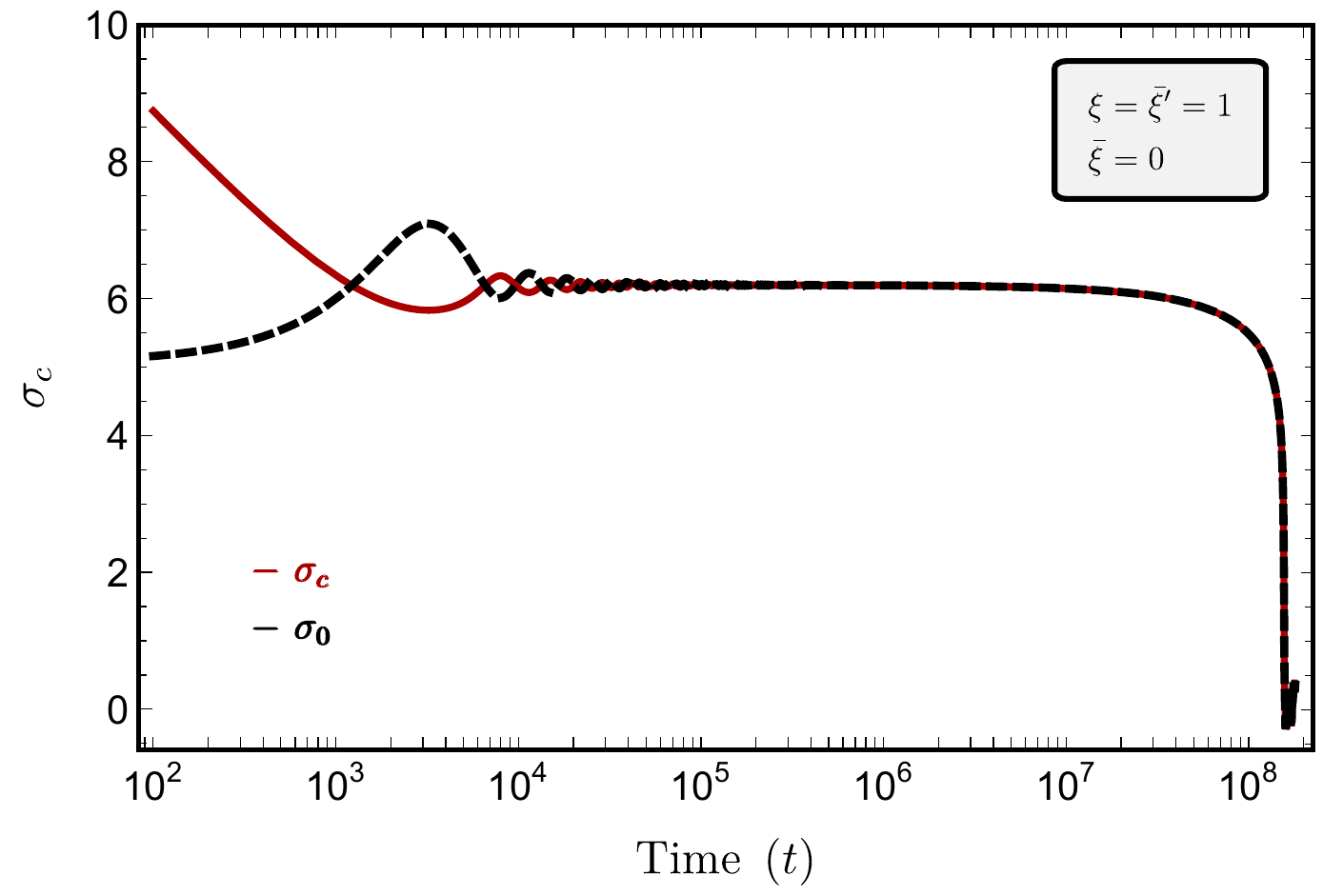}
\includegraphics[scale=0.546]{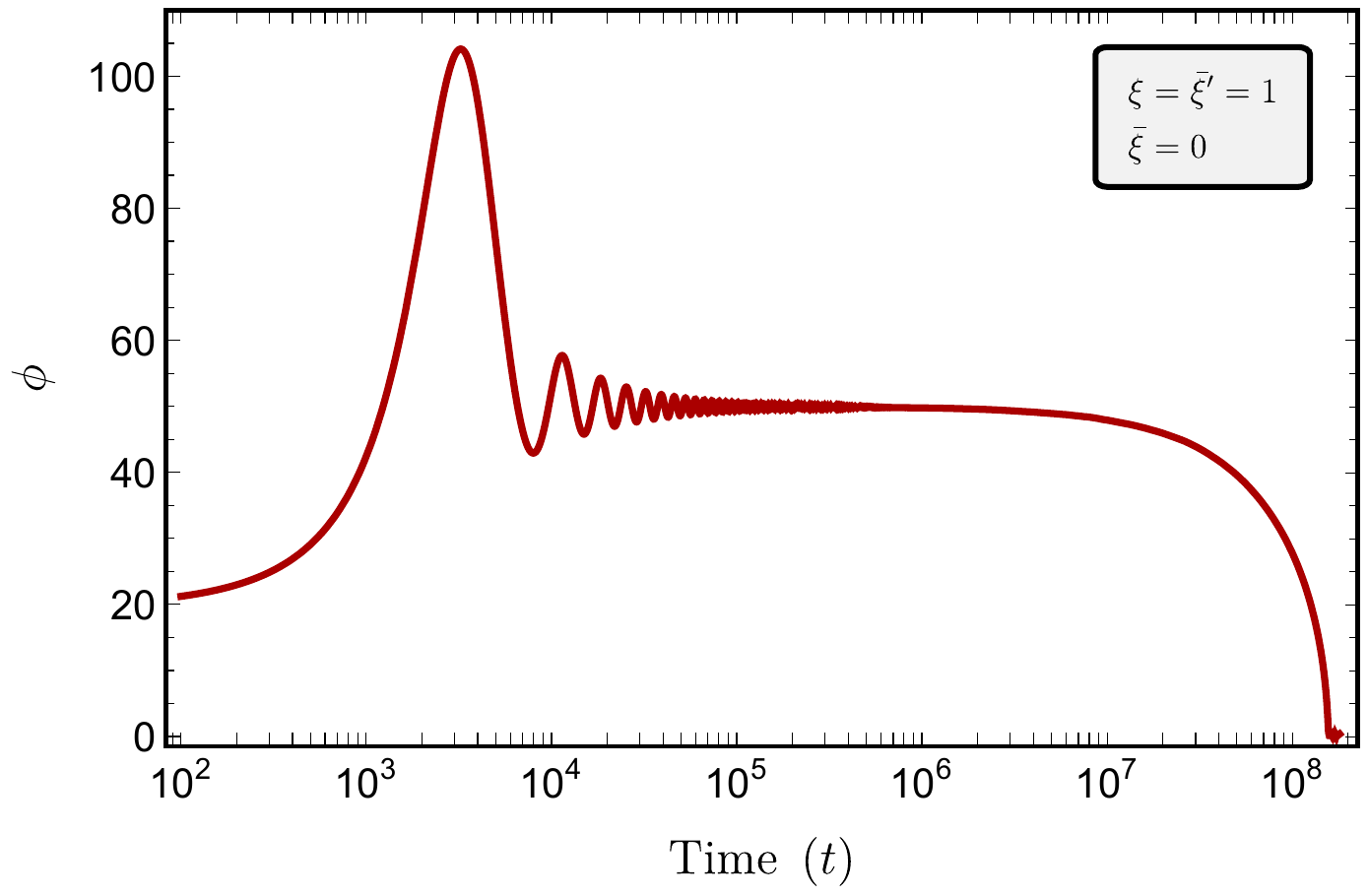}
\caption{The evolution of the scalar fields $\s_c$ (left) and $\phi$ (right). Both fields start from the initial field value $20$ with zero velocities. The black dashed line in the left plot corresponds to the minimum direction $\s_0(\phi)$. The parameters are $\g=10^6$, $\xi=\bar{\xi}'=1$ and $\bar{\xi}=0$.  }
\label{fig:two_field_prob}
\end{center}
\end{figure}

We have considered the system of differential equations for $\phi(t)$ and $\sigma_c(t)$ and solved it numerically. We start with indicative initial field values $\sigma_c(0)=\phi(0)=20$ and zero velocities $\dot{\sigma}_c(0)=\dot{\phi}(0)=0$. The parameters are chosen at the characteristic values $\g=10^6$, $\xi=\bar{\xi}'=1$ and $\bar{\xi}=0$. The results are shown in figure~\ref{fig:two_field_prob}. It is evident that when $\phi$ reaches the inflationary plateau, the $\sigma_c$ has already fallen into the valley defined by the minimum line~\eqref{eq:sigma_0}. This can be also understood by comparing the relevant mass of the $\sigma_c-$field with the inflationary Hubble scale. After a numerical calculation we obtain that $M_{\sigma_c}^2(\phi)/H^2 \sim 10^4$ (for the benchmark case of figure~\ref{fig:two_field_prob}) at times $t \gtrsim 10^5$, which reinforces further the conclusion that the model is downgraded to a \textit{single-field} inflationary model.

Therefore, for the inflationary period it would be sufficient to study the {\textit{single-field problem}} described by $\phi,\sigma_0(\phi)$. Then, the system is described by the single field action
\be 
{\cal{S}}=\int\,{\rm d}^4x\,\sqrt{-g}\left\{\frac{1}{2}R-\frac{1}{2}\bar{K}(\phi)(\nabla\phi)^2+\frac{1}{4}L(\phi)(\nabla\phi)^4-U(\phi)\,\right\},
\label{eq:act1f}
\ee
where
\ba
\bar{K}(\phi) &=& \frac{f(\phi)}{f^2(\phi)+4\gamma V(\phi)}+\left(\frac{12}{1+\frac{4g(\phi)^2f^2(\phi)}{[f^2(\phi)+4\gamma V(\phi)]^2}}\right)\left(\frac{g'(\phi)f(\phi)+g(\phi)f'(\phi)}{f^2(\phi)+4\gamma V(\phi)}\right. \nonumber
\\  & &\left.-\frac{g(\phi)f(\phi)}{[f^2(\phi)+4\gamma V(\phi)]^2}\left(2f'(\phi)f(\phi)+4\gamma V'(\phi)\right)\right)^2\,, \nonumber
\\ L(\phi) &=&\frac{\gamma}{\gamma g^2(\phi)/\delta+f^2(\phi)+4\gamma V(\phi)}\,, \nonumber
\\ U(\phi) & =& \frac{V(\phi)}{f^2(\phi)+4\gamma V(\phi)}\,.
\ea
Note that both kinetic functions $\bar{K}(\phi)$ and $L(\phi)$ are positive definite.

Considering an FRW background, we obtain from~\eqref{eq:act1f} the following set of equations of motion
\ba
&&\left(\bar{K}(\phi)+3L(\phi)\dot{\phi}^2\right)\ddot{\phi} + 3H\left(\bar{K}(\phi)+L(\phi)\dot{\phi}^2\right)\dot{\phi}+\frac{1}{2}\bar{K}'(\phi)\dot{\phi}^2+\frac{3}{4}L'(\phi)\dot{\phi}^{4}+U'(\phi)=0\,, \nonumber
\\ && H^2 = \frac{\rho}{3}\quad \text{and} \quad  \dot{H}=-\frac{1}{2}\left(\rho+p\right)\,,
\label{eq:eoms}
\ea
where the energy density and pressure are given by
\be
\rho=\frac{1}{2}\bar{K}(\phi)\dot{\phi}^2+\frac{3}{4}L(\phi)\dot{\phi}^4+U(\phi) \quad \text{and} \quad p=\frac{1}{2}\bar{K}(\phi)\dot{\phi}^2+\frac{1}{4}L(\phi)\dot{\phi}^4-U(\phi)\,. 
\ee
Although we have displayed the single-field equations of motion for general $f(\phi),\,g(\phi),\,V(\phi)$, in the analysis of inflationary observables that will follow we have focused on $f(\phi)=1+\xi\phi^2$, $g(\phi)=\bar{\xi}\phi+\bar{\xi}'\phi^3$ and $V(\phi)=\lambda\phi^4/4$.

Due to the appearance of the higher order kinetic term in the equation of motion the speed of sound in general deviates from unity. This can be easily seen from the form of the speed of sound which is given by
\be
c_s^2 = \frac{1+L(\phi)\dot{\phi}^2/\bar{K}(\phi)}{1+3L(\phi)\dot{\phi}^2/\bar{K}(\phi)}\,.
\ee
Nevertheless, the deviation from unity turns out to be quite small. 

\subsection{Inflationary observables}
\label{sec:observables}
As it is quite complicated to find analytical expressions for the inflationary observables, we will solve the equation of motion~\eqref{eq:eoms} numerically to calculate them. To be as precise as possible the calculation of the amplitude of the scalar power spectrum $A_s$, the spectral index $n_s$, and the tensor-to-scalar ratio $r$, is done in the framework of the adiabatic approximation as described in~\cite{Martin:2013uma}. This method is suitable for theories with varying speed of sound and resembles to the well known WKB method. In our model even though the speed of sound is not constant, it is very close to unity throughout the full inflationary period, resulting to insignificant corrections to the inflationary observables.

Following~\cite{Martin:2013uma}, in order to calculate the observables we will use the Hubble flow functions given by
\be
\label{eq:hff}
\epsilon_1 = -\frac{\dot{H}}{H^2}\,, \quad \epsilon_2 = \frac{\dot{\epsilon_1}}{\epsilon_1 H}\,, \quad s_1 = \frac{\dot{c_s}}{c_s H}\,.
\ee
The scalar ($\mathcal{P}_\zeta$) and tensor ($\mathcal{P}_h$) power spectra can be expanded around the arbitrary pivot scale $k_\star= a_\star H_\star/c_s^\star$, that exited the sound horizon at some time $t^\star$. keeping the first order terms in
the Hubble flow functions~\eqref{eq:hff}, we obtain
\ba
\mathcal{P}_\zeta(k) &=& \frac{H_\star^2}{8\pi^2 \epsilon_1^\star c_s^\star}\left(1-2(D+1)\epsilon_1^\star -D\epsilon_2^\star -(2+D)s_1^\star -(2\epsilon_1^\star +\epsilon_2^\star +s_1^\star) \ln \frac{k}{k_\star}\right)\,,
\\ \mathcal{P}_h(k) &=& \frac{2H_\star^2}{\pi^2} \left(1 -2(D+1-\ln c_s^\star)\epsilon_1^\star -2\epsilon_1^\star \ln \frac{k}{k_\star} \right)\,,
\ea
where the constant $D$ is given by $D=7/19 -\ln3$.
The corresponding amplitudes, for scalar and tensor perturbations, are given by,
\be
A_s^\star = \mathcal{P}_\zeta(k_\star) \quad \text{and} \quad A_t^\star = \mathcal{P}_h(k_\star)\,.
\ee
The Planck 2018 data~\cite{Akrami:2018odb}, yield a value for the amplitude of the scalar power spectrum $A_s^\star=(2.10\pm0.03) \times 10^{-9}$, at the pivot scale $k_\star=0.05\, {\rm Mpc^{-1}}$.

The tensor-to-scalar ratio ($r$) and the spectral index of the scalar power spectrum ($n_s$) are given by
\ba
r &=& \frac{A_t^\star}{A_s^\star} = 16\epsilon_1^\star c_s^\star\left(1+2\epsilon_1^\star\ln c_s^\star  +D \epsilon_2^\star +(2+D)s_1^\star \right)\,,
\\  n_s &=&1-2\epsilon_1^\star -\epsilon_2^\star -s_1^\star\,.
\ea

The recent release of the BICEP/Keck~\cite{BICEP:2021xfz} data, imposes the bound $r_{0.05} < 0.036$ at the $95\%$ C.L. where the subscript denotes the pivot scale in $ {\rm Mpc^{-1}}$. Furthermore, the combination of WMAP,
Planck and BICEP/Keck data constrains the spectral index to the range $0.958<n_s<0.975$ again at the $95\%$ C.L.  for $r=0.004$.

After the end of inflation, the Universe enters to the radiation-dominated era through a reheating phase, which may last some $e-$folds or can be instantaneous. Following~\cite{Liddle:2003as}, we can compute the number of $e-$folds during  
inflation after the comoving scale, $k_\star= a_\star H_\star/c_s^\star$, crosses the sound horizon by defining $N_\star=\ln\left(\frac{a_{\rm end}}{a_\star} \right)$, where $a_{\rm end /{}_\star}$ denote the scale factor at the end of inflation and at the scale $k_\star$ respectively. Analyzing the  scale factor as
\be
c_s^\star k_\star = \frac{a_\star}{a_{\rm end}} \frac{a_{\rm end}}{a_{\rm reh}} \frac{a_{\rm reh}}{a_{0}} a_o H_\star
\ee
and assuming instantaneous reheating ($a_{\rm end}=a_{\rm reh}$) and entropy conservation, we obtain
\be
\label{eq:N_star}
N_\star=66.89-\ln c_s^\star -\ln\left( \frac{k_\star}{a_0 H_0}\right) +\frac{1}{2} \ln\left( \frac{3 H_\star^2}{\rho_{\rm end}^{1/2}}\right) -\frac{1}{12}\ln g^s_{\rm reh}.
\ee
\begin{figure}[t!]  
\begin{center}
\includegraphics[scale=0.8]{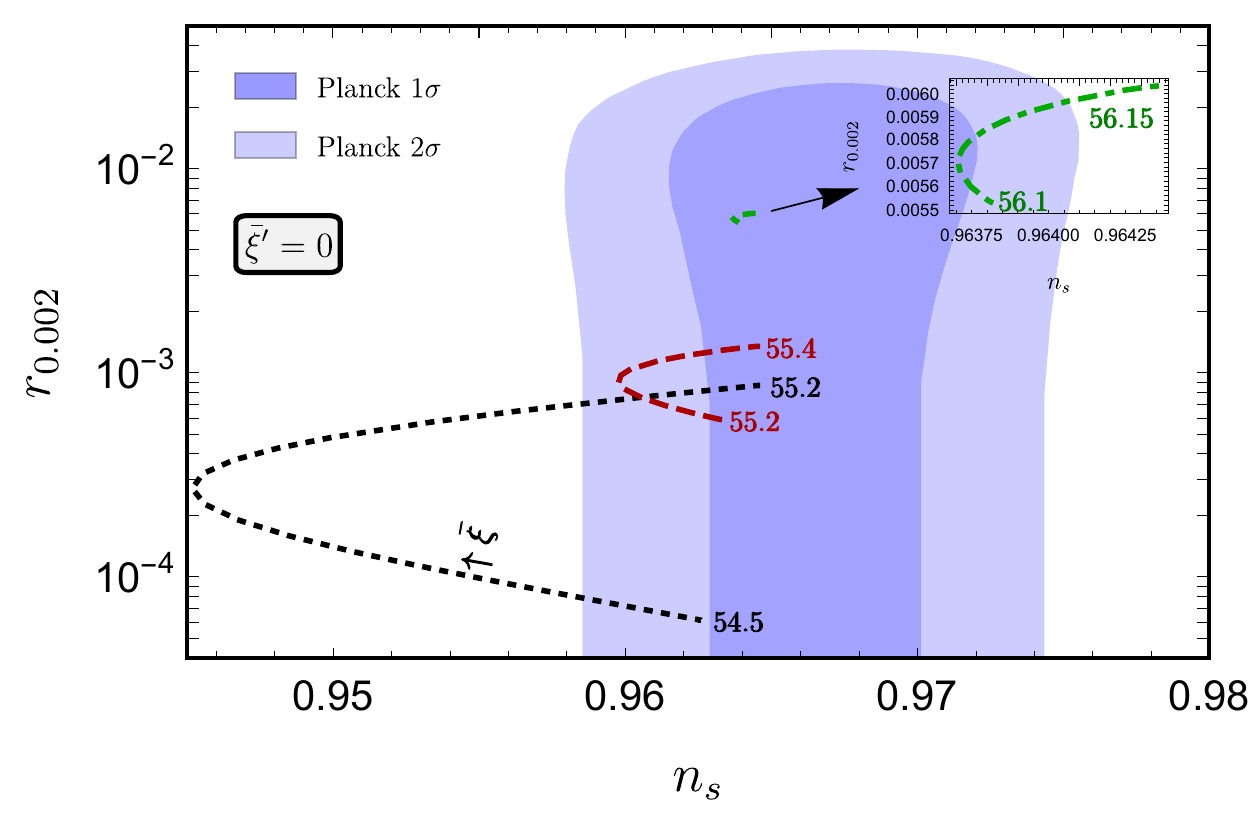}
\caption{Predictions of the model using pivot scales $0.05\, {\rm Mpc^{-1}}$ for $n_s$ and $0.002\, {\rm Mpc^{-1}}$ for $r$.  Shaded regions are the allowed parameter regions
at $68\%$ and $95\%$ confidence coming from the latest
combination of Planck, BICEP/Keck and BAO data~\cite{Akrami:2018odb,BICEP:2021xfz}. The values of the parameters are $\bar{\xi}'=0$ and $\g=10^6$, while $\xi=0.1$ (green dashed-dotted line), $\xi=1$ (red dashed line) and $\xi=10$ (black dotted line). The parameter $\bar{\xi}$ varies from $10^{-3}$ to $10^3$ in each curve in a clockwise direction indicated by the arrow. The small numbers at the edges of the curves indicate the number of $e-$folds $N_{0.05}$ calculated by~\eqref{eq:N_star}, for the extreme values of the parameter $\bar{\xi}$.} 
\label{fig:ns_r_1}
\end{center}
\end{figure}
The subscripts ``reh" and ``0" denote evaluations at the end of reheating and  present epoch respectively, while $g^s_{\rm reh}$ are the entropy density degrees of freedom being $106.75$ assuming the Standard Model particle content and temperatures $\sim 1\, \TeV$ or higher. Having in mind that $V_{\rm eff}^\star \simeq \frac{3\pi^2}{2}A_s^\star r_\star$, $\rho_{\rm end}\simeq \frac{3}{2} V_{\rm eff}^{\rm end}$,~\footnote{This equality is exact if there are no higher order kinetic terms. In~\cite{Gialamas:2019nly} the higher-order kinetic terms have been taken into account, but as it is shown there, only an insignificant correction arises.} and omitting the speed of sound contribution we obtain for the pivot scale $k_\star=0.05\, {\rm Mpc^{-1}}$ that

\begin{table}[h!]
\centering
\caption{Predictions of the model for the benchmark points that correspond to the maximum values of the spectral index $n_s$ of figure~\ref{fig:ns_r_1}.}
\begin{tabular}{ccccccc}
\hline  \\[-4.2mm]\hline\\[-2.2mm]
$\xi$& $\bar{\xi} $& ${n_s}_{0.05}$ & $r_{0.05}$  & $N_{0.05}$ & $r_{0.002}$ & $N_{0.002}$ \\
\hline  \\[-3.2mm]
$0.1$ & $3 $ & $0.9644$&  $6.72\times 10^{-3} $ & $56.15$ & $6.04\times 10^{-3}$ & $59.37$ \\
$1$ & $30$ &$0.9645$ & $1.50\times 10^{-3}$  & $55.39$ & $1.35\times 10^{-3}$ & $58.61$\\
$10$ & $300 $& $0.9645$ & $9.66\times 10^{-4}$  & $55.20$ & $8.66\times 10^{-4}$ & $58.42$   \\
 \hline \hline
\end{tabular}
\label{tbl_1}
\end{table}

\be
N_{0.05} \simeq 55.8 +\frac{1}{4}\ln \frac{r}{0.036} -\frac{1}{4}\ln \frac{ V_{\rm eff}^{\rm end}}{V_{\rm eff}^{0.05}}.
\ee
At the pivot scale $k_\star=0.002\, {\rm Mpc^{-1}}$, in which the tensor-to-scalar ratio is calculated in some of our figures, the numerical coefficient increases to the value $\sim 59.1$.
\begin{figure}[t!]  
\begin{center}
\includegraphics[scale=0.71]{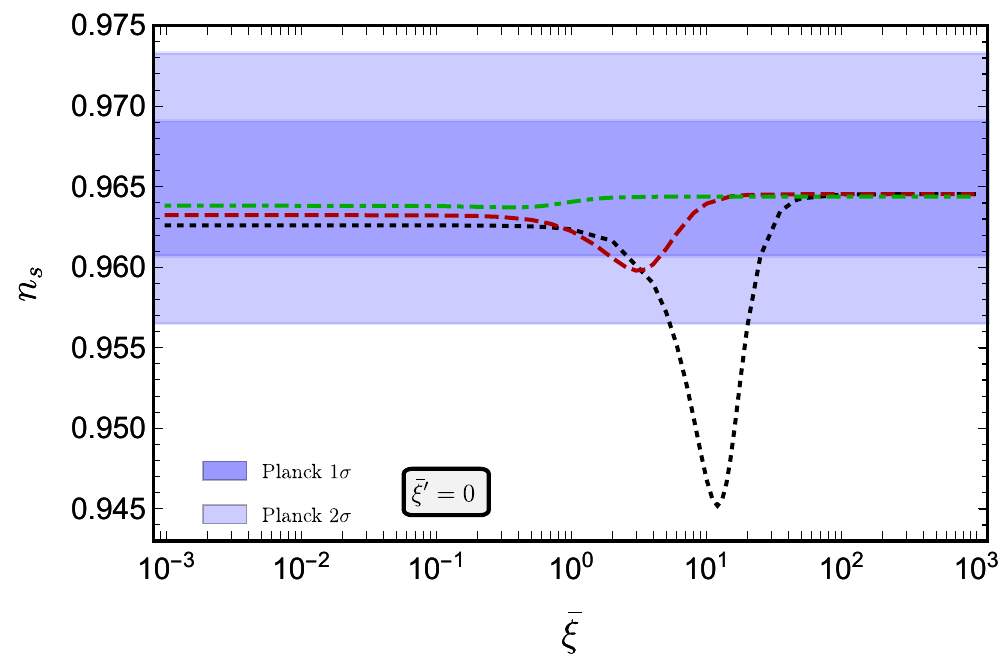}
\includegraphics[scale=0.515]{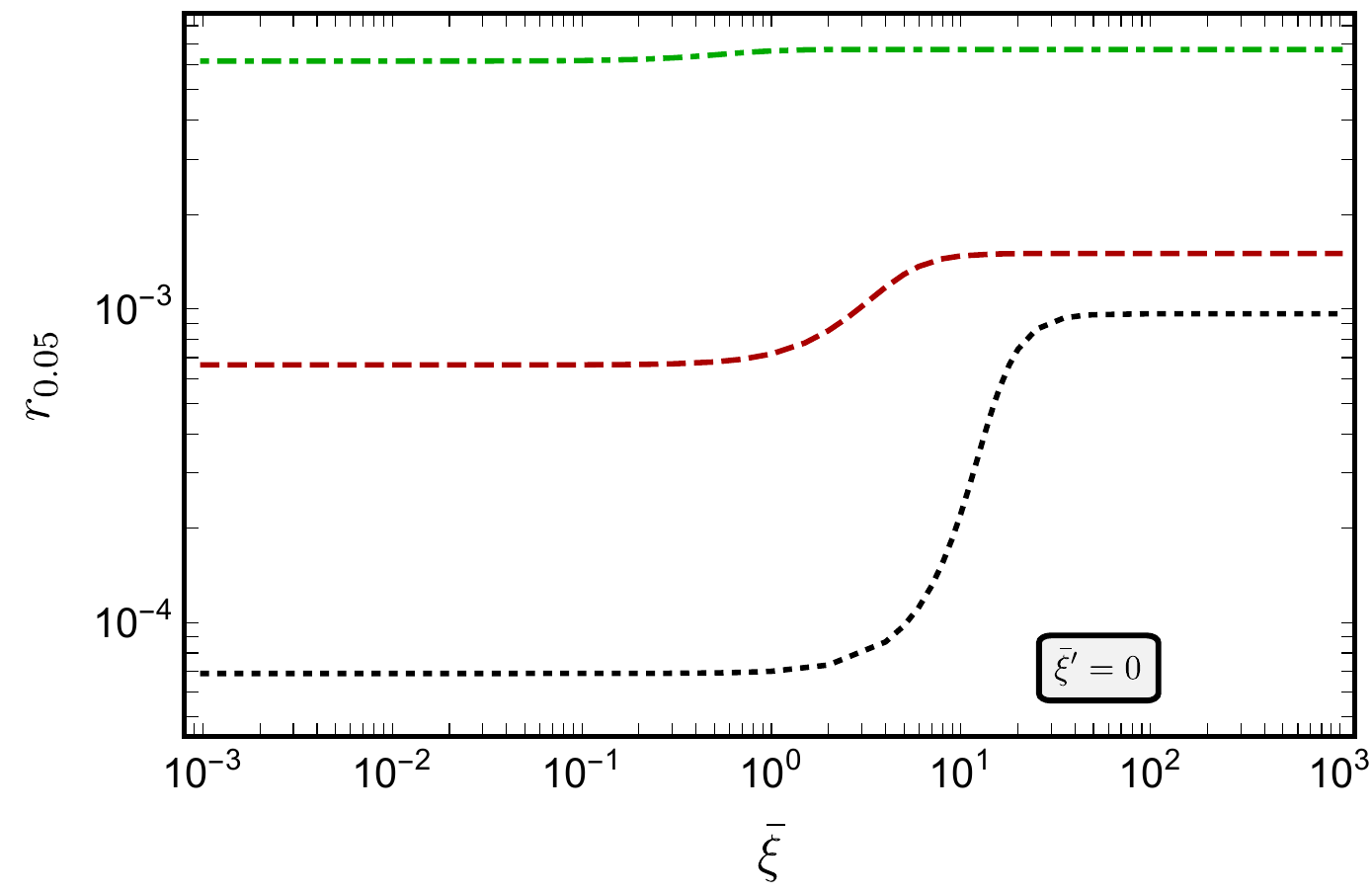}
\caption{Predictions of the model as a function of the parameter $\bar{\xi}$ using the pivot scale $0.05\, {\rm Mpc^{-1}}$ for both $n_s$ and  $r$.  Shaded regions in the left plot are the allowed parameter regions, $n_s=0.9649\pm 0.0042$, at $68\%$ and $95\%$ confidence coming from the Planck data~\cite{Akrami:2018odb}. In the right plot the whole region is allowed, since $r_{0.05}< 0.036$ at $95\%$ confidence~\cite{BICEP:2021xfz}. The values of the parameters and the coloring are the same with that in figure~\ref{fig:ns_r_1}.}
\label{fig:ns_r_xibar_1}
\end{center}
\end{figure}

Let us now turn to the analysis of the parametric space of the model under consideration with respect to its predictions for the cosmological observables. For the parameter $\xi$ we choose to start from the value $\xi=0.1$, since as discussed in~\cite{Gialamas:2019nly} in the ``simple" Palatini-$\mathcal{R}^2$ model for values $\xi \gtrsim 0.1$ the spectral index can be within observational limits. It is also acceptable to use lower values, but at the price of reducing the range of acceptable e-folds considerably. For vanishing $g(\phi)$ the spectral index and the tensor-to-scalar ratio are given approximately by
\be
\label{eq:ns_r+PalR2}
n_s \simeq 1-\frac{2}{N_\star}-\frac{1}{8\xi N_\star^2}\quad \text{and} \quad r \simeq \frac{1}{N_\star^2}\frac{2\xi}{\xi^2+\g\l}\,. 
\ee

In our case both of them get modified, but their analytic expressions are not available since the functions involved are too complicated. We will examine two limiting cases for the function $g(\phi)$ for the sake of  better understanding of the behaviour of the observables.

In figure~\ref{fig:ns_r_1} we explore the case in which the linear part of the non-minimal coupling function $g(\phi)$ of Eq.~\eqref{eq:g_func} dominates, taking the parameter $\bar{\xi}'$ to be zero. In this figure we show the predictions of the model in the $\s_0(\phi)$ direction using pivot scales $0.05\, {\rm Mpc^{-1}}$ for $n_s$ and $0.002\,{\rm Mpc^{-1}}$ for $r$. For each given set of values for the parameters, we have employed Eq.~\eqref{eq:N_star} to obtain the number of $e$-folds that complies with the constraints from reheating, while the value of quartic coupling $\lambda$ is fixed by the observed value of the amplitude of the scalar power spectrum, $A_s^\star=2.1 \times 10^{-9}$, at the pivot scale $k_\star=0.05\, {\rm Mpc^{-1}}$. The parameter $\g$ is fixed to the value $10^6$, while $\xi=0.1$ (green dashed-dotted line), $\xi=1$ (red dashed line) and $\xi=10$ (black dotted line). 
The parameter $\bar{\xi}$ varies from $10^{-3}$ to $10^3$ in each curve in a clockwise direction as indicated in the figure. As the parameter $\bar{\xi}$ gets larger the tensor-to-scalar ratio $r$ increases. The spectral index $n_s$ decreases up to some value of $\bar{\xi}$ beyond which it starts to increase. A notable remark is that the predictions for $n_s$ and $r$ ``freeze" to some values (their largest values) without being affected by a further increasing of the parameter $\bar{\xi}$. This ``freezing" occurs at the values of $\bar{\xi} \simeq 3, 30$ and $300$ for $\xi=0.1, 1$ and $10$ respectively, see also Table~\ref{tbl_1}. In figure~\ref{fig:ns_r_xibar_1} we also present the observables versus  the parameter $\bar{\xi}$, for better understanding of the analysis above.
\begin{figure}[t!]  
\begin{center}
\includegraphics[scale=0.8]{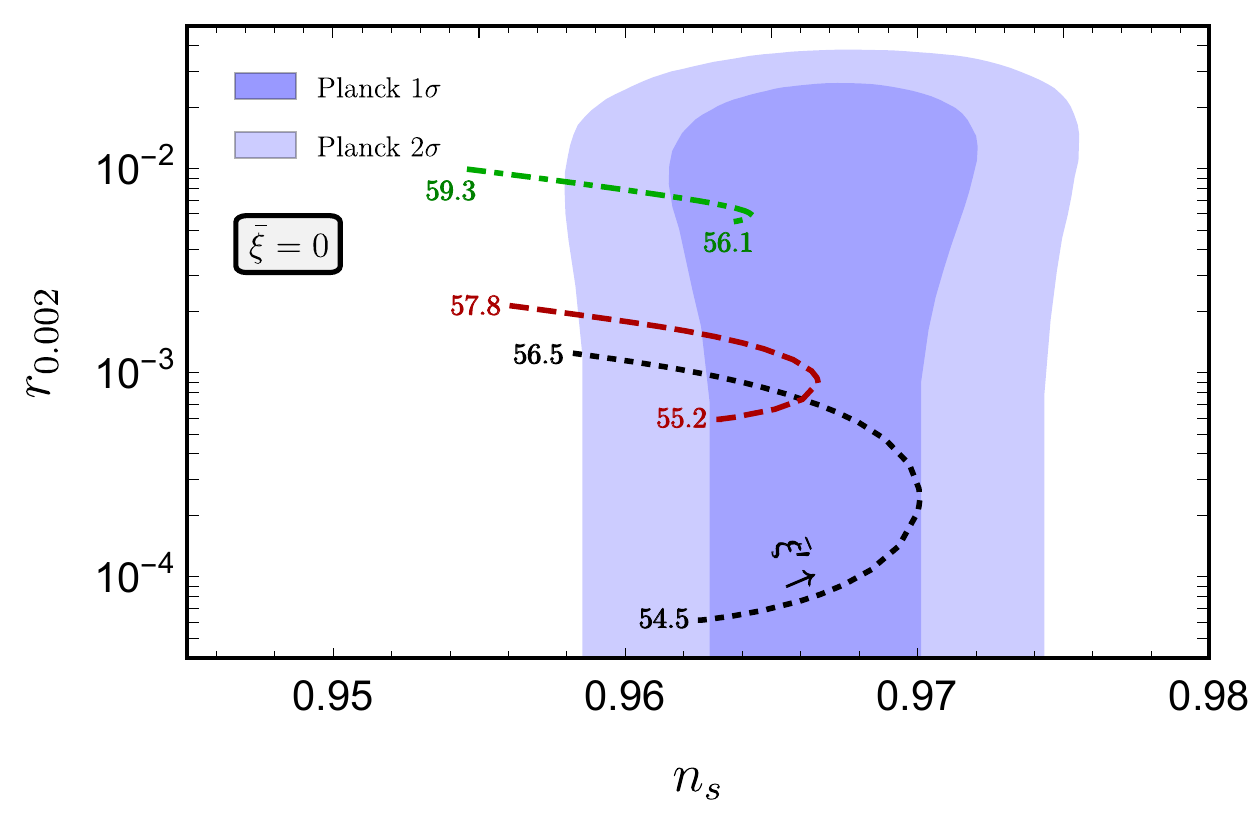}
\caption{Predictions of the model using pivot scales $0.05\, {\rm Mpc^{-1}}$ for $n_s$ and $0.002\, {\rm Mpc^{-1}}$ for $r$.  Shaded regions are the allowed parameter regions
at $68\%$ and $95\%$ confidence coming from the latest
combination of Planck, BICEP/Keck and BAO data~\cite{Akrami:2018odb,BICEP:2021xfz}. The values of the parameters are $\bar{\xi}=0$ and $\g=10^6$, while $\xi=0.1$ (green dashed-dotted line), $\xi=1$ (red dashed line) and $\xi=10$ (black dotted line). The parameter $\bar{\xi}'$ varies from $10^{-3}$ to $10^3$ in each curve in a counterclockwise direction indicated by the arrow. The small numbers at the edges of the curves indicate the number of $e-$folds $N_{0.05}$ calculated by~\eqref{eq:N_star}, for the extreme values of the parameter $\bar{\xi}'$.}
\label{fig:ns_r_2}
\end{center}
\end{figure}

\begin{figure}[t!]  
\begin{center}
\includegraphics[scale=0.72]{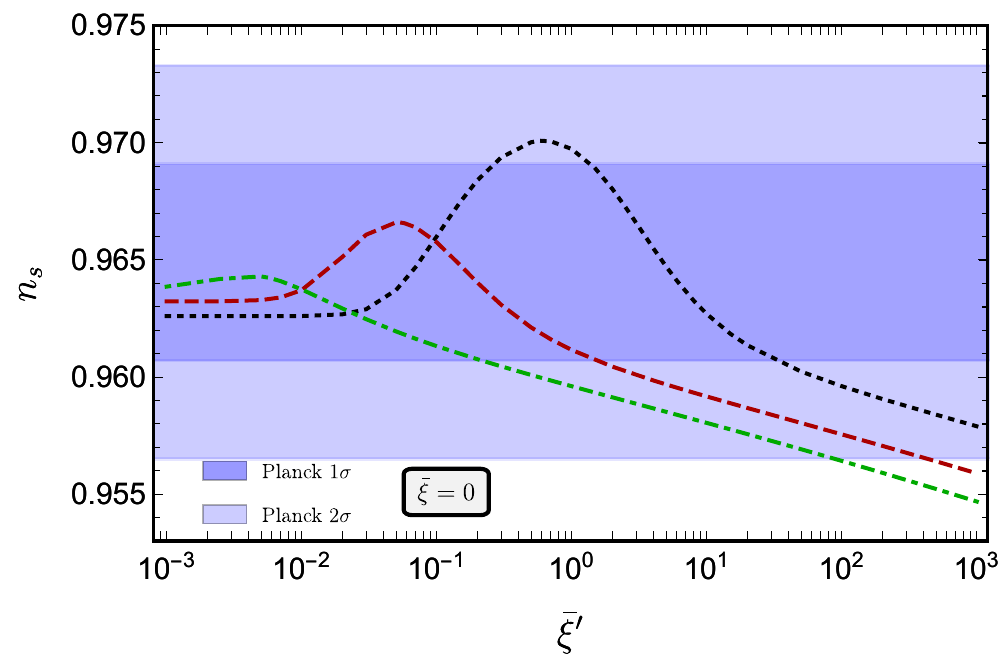}
\includegraphics[scale=0.515]{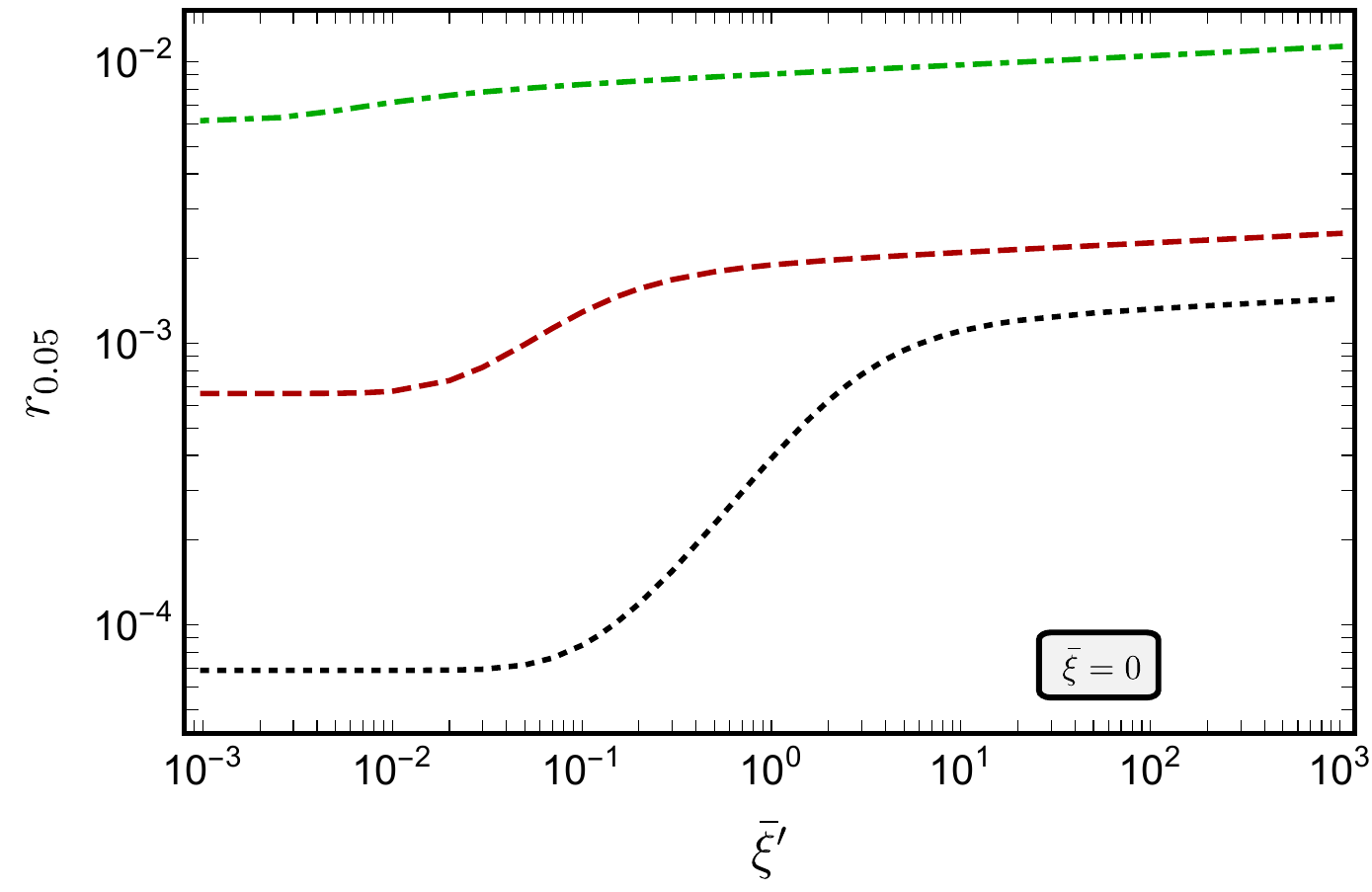}
\caption{Predictions of the model as a function of the parameter $\bar{\xi}'$ using the pivot scale $0.05\, {\rm Mpc^{-1}}$ for both $n_s$ and  $r$.  Shaded regions in the left plot are the allowed parameter regions, $n_s=0.9649\pm 0.0042$, at $68\%$ and $95\%$ confidence coming from the Planck data~\cite{Akrami:2018odb}. In the right plot the whole region is allowed, since $r_{0.05}< 0.036$ at $95\%$ confidence~\cite{BICEP:2021xfz}. The values of the parameters and the coloring are the same with that in figure~\ref{fig:ns_r_2}.}
\label{fig:ns_r_xibar_2}
\end{center}
\end{figure}

In Figs.~\ref{fig:ns_r_2} and~\ref{fig:ns_r_xibar_2} we explore the case in which the cubic  part of the non-minimal coupling function $g(\phi)$ of Eq.~\eqref{eq:g_func} dominates, \textit{i.e.} we choose $\bar{\xi}=0$. In this case the tensor-to-scalar ratio $r$ is constantly increasing as $\bar{\xi}'$ gets larger (from $10^{-3}$ to $10^3$), while the spectral index $n_s$ initially increases, gets its maximum value and then starts to decrease and go well outside of the observational acceptable bounds~\cite{Akrami:2018odb,BICEP:2021xfz}. The datasets that correspond to the maximum value of $n_s$ are given in Table~\ref{tbl_2}. The values of the rest parameters and the
coloring of Figs.~\ref{fig:ns_r_2} and~\ref{fig:ns_r_xibar_2} are the same with that in  Figs.~\ref{fig:ns_r_1} and~\ref{fig:ns_r_xibar_1}.

\begin{table}[h!]
\centering
\caption{Predictions of the model for the benchmark points that correspond to the maximum values of the spectral index $n_s$ of figure~\ref{fig:ns_r_2}.}
\begin{tabular}{ccccccc}
\hline  \\[-4.2mm]\hline\\[-2.2mm]
$\xi$& $\bar{\xi}' $& ${n_s}_{0.05}$ & $r_{0.05}$  & $N_{0.05}$ & $r_{0.002}$ & $N_{0.002}$ \\
\hline  \\[-3.2mm]
$0.1$ & $5\times 10^{-3} $ & $0.9643$&  $6.70\times 10^{-3} $ & $56.13$ & $6.01\times 10^{-3}$ & $59.35$ \\
$1$ & $5\times 10^{-2}$ &$0.9666$ & $9.92\times 10^{-4}$  & $55.28$ & $8.95\times 10^{-4}$ & $58.50$\\
$10$ & $6\times 10^{-1} $& $0.9701$ & $2.63\times 10^{-4}$  & $54.81$ & $2.40\times 10^{-4}$ & $58.02$   \\
 \hline \hline
\end{tabular}
\label{tbl_2}
\end{table}
 
In all of the figures presented the parameter $\gamma$ has the value $10^6$, but for any value lower than $\sim 10^7$ the predictions will remain unaffected. This can be easily understood if we have a look at the large field limits of the effective potential and the kinetic term $\bar{K}(\phi)$ (see Eq.~\eqref{EFFLANG}). For both of them the limits are functions of the combination $\xi^2/\l +\g$, which means that if $\xi^2/\l \gg \g$ the inflationary predictions are controlled by $\xi^2/\l$. In the parameter space under consideration the ratio $\xi^2/\l$ is always larger than $\sim 7\times 10^8$. In particular, in Figs.~\ref{fig:ns_r_2}-\ref{fig:ns_r_xibar_2} varies from $1.2\times 10^9 - 6.6\times  10^8$ (for $\xi=0.1$), $1.2\times 10^{10} - 3.2\times 10^9$ (for $\xi=1$) and $1.2\times 10^{11} - 5.5\times 10^9$ (for $\xi=10$). Similar values are noticed also in Figs.~\ref{fig:ns_r_1}-\ref{fig:ns_r_xibar_1}. At this point, we would like to mention that we do not consider larger values of $\g$ that would drive us to the regime $\xi^2/\l \ll \g$, since the tensor-to-scalar ratio would be even smaller. The sensitivity in the measurement of the tensor-to-scalar ratio by the next-generation CMB satellites~\cite{Matsumura2016,Kogut_2011,Hanany:2019lle} is expected to be $\d r\sim 10^{-4}$, so smaller values for $r$ are only subject of academic significance. The general picture is that the presence of the pseudoscalar and its coupling to the inflaton through the term $g(\phi)\widetilde{\cal{R}}$ has as a result an increase of the tensor-to-scalar ratio, in comparison to the Palatini-${\cal{R}}^2$ model, while the spectral index can either decrease or increase depending on the rest of the parameters. 
In the whole range of the aforementioned parameters the number of $e$-folds is calculated to be $N_{0.05}\gtrsim 55$, under the assumption of instantaneous reheating. A smaller number of $e$-folds can arise for larger values of $\xi$, nevertheless, this would also lead to a smaller value for the tensor-to-scalar ratio.

\section{Conclusions}
\label{sec:conclusions}

In the present article we considered the framework of \textit{Metric-Affine} theories of gravity in which the metric and the connection are independent variables. We extended the Einstein-Hilbert action to include the parity-odd Holst invariant $\widetilde{\cal{R}}$ as well as quadratic terms of it and of the Ricci scalar, expected to be generated by quantum corrections. This context gives rise to a new dynamical pseudoscalar degree of freedom. Furthermore, we considered the coupling of this system to a scalar field $\phi$ through the non-minimal terms $f(\phi){\cal{R}}$ and $g(\phi)\widetilde{\cal{R}}$, the latter corresponding to a coupling of $\phi$ with the new pseudoscalar. The potential of the resulting Einstein-frame action of the two fields $\phi$ and $\sigma$ has a minimum line along $\sigma_0(\phi)$. Considering this model in an FRW background we found that the pseudoscalar $\sigma$ falls quickly into the valley defined by the minimum line $\sigma_0(\phi)$, this transition being supported by the numerical solution of the two-field system of equations of motion. Thus, the model is reduced to an effectively single-field model.  Furthermore, along this minimum line the potential takes the form of the Palatini-${\cal{R}}^2$ potential with its characteristic inflationary plateau.

We have proceeded to study the resulting single-field inflationary model making a choice of the coupling functions $f(\phi)=1+\xi\phi^2$ and $g(\phi)=\bar{\xi}\phi+\bar{\xi}'\phi^3$. A numerical calculation of the inflationary observables was undertaken in the framework of the adiabatic approximation \cite{Martin:2013uma}. The parametric space was constrained by the latest available bounds set by~\cite{Akrami:2018odb,BICEP:2021xfz}. We have found 
that our model complies with observations for a wide range of parameters. More precisely taking the parameter $\xi$ to be $\geq 0.1$, the parameters $\bar{\xi}$ and $\bar{\xi}'$ run in the range $10^{-3}-10^{3}$. 
 We examined two limiting cases for the function $g(\phi)$ in order to better understand the behaviour of the observables: the linear case ($\bar{\xi}'=0$) and the qubic case ($\bar{\xi}=0$). In the linear case  the predictions for $n_s$ and $r$ ``freeze" to their largest values without being affected by a further increase of the parameter $\bar{\xi}$. This ``freezing" occurs at the values of $\bar{\xi} \simeq 3, 30$ and $300$ for $\xi=0.1, 1$ and $10$ respectively. In the cubic case after the initial increase of the spectral index, once it gets its maximum value, it starts to decrease and go well outside of the observational acceptable bounds as $\bar{\xi}' $ gets larger. In general we found that the coupling $g(\phi)$ has as a result an increase of the tensor-to-scalar ratio, in comparison to the Palatini-${\cal{R}}^2$ model, while the spectral index can either decrease or increase depending on the position in parametric space.
The coefficient of the $ \widetilde{\mathcal{R}}^2$ term does not modify the inflationary observables, while the one of  the $ \mathcal{R}^2$ term can have a significant result only if it is larger than $10^7$, which case is not considered here. 
Finally, under the assumption of instantaneous reheating, in the whole range of the aforementioned parameters the number of $e$-folds measured from the end of inflation to the pivot scale $k_\star=0.05\, {\rm Mpc^{-1}}$ is $N_{0.05}\gtrsim 55$. 

Summarizing, we have considered a general quadratic \textit{Metric-Affine} theory, featuring an extra dynamical degree of freedom,  and coupled it non-minimally to a scalar field. We studied inflation in the resulting two-field model and found that it effectively reduces to a single-field model, with a potential of the Palatini-${\cal{R}}^2$ form with its characteristic inflationary plateau and a modified kinetic term. We find that the inflationary predictions of this model fall within the latest observational bounds for a wide range of parameters. Furthermore, it allows for an increase in the tensor-to-scalar ratio.

\acknowledgments

The work of IDG was supported by the Estonian Research Council grant SJD18.
KT would like to thank the CERN Theoretical Physics Department and the Laboratoire de Physique de l’Ecole Normale Superieure, where part of this work was done, for their hospitality.

\vspace{1 cm}
\bibliography{Metric_Affine_Inflation_G_T}{}
\end{document}